\documentclass[aps, prl, twocolumn, showpacs,nofootinbib,superscriptaddress]{revtex4} 
\usepackage{graphicx}
\usepackage{epsfig} 
\usepackage{amsmath} 
\usepackage{amsfonts}
\usepackage{amssymb} 
\usepackage{hyperref}
\usepackage{color} 

\usepackage{soul} 

\newcommand{\ket}[1]{|{#1}\rangle} 
\newcommand{\bra}[1]{\langle{#1}|}
\newcommand{\bracket}[2]{\langle{#1}|{#2}\rangle}
\newcommand{\ketbra}[2]{|{#1}\rangle\langle{#2}|}

\newcommand{\beq}{\begin{equation}}
\newcommand{\eeq}{\end{equation}} 
\newcommand{\bea}{\begin{eqnarray}}
\newcommand{\eea}{\end{eqnarray}}
\newcommand{\tr}{\operatorname{Tr}}

\newcommand{\sys}{\mathcal{S}}
\newcommand{\aux}{\mathcal{A}}
\newcommand{\norminf}[1]{\left|\left|#1\right|\right|_{\infty}}

\begin{document}

\author{Augusto J. Roncaglia}
\affiliation{Departamento de F\'\i sica, FCEyN, UBA, Ciudad
Universitaria Pabell\'on 1, 1428 Buenos Aires, Argentina}
\affiliation{IFIBA CONICET, FCEyN, UBA, Ciudad Universitaria
Pabell\'on 1, 1428 Buenos Aires, Argentina} 

\author{Federico Cerisola} 
\affiliation{Departamento de F\'\i sica, FCEyN, UBA, Ciudad 
Universitaria Pabell\'on 1, 1428 Buenos Aires, Argentina}

\author{Juan Pablo Paz}
\affiliation{Departamento de F\'\i sica, FCEyN, UBA, Ciudad
Universitaria Pabell\'on 1, 1428 Buenos Aires, Argentina}
\affiliation{IFIBA CONICET, FCEyN, UBA, Ciudad Universitaria
Pabell\'on 1, 1428 Buenos Aires, Argentina}

\title{Work measurement as a generalized quantum measurement}


\begin{abstract} 
We present a new method to measure the work $w$ 
performed on a driven quantum system and to 
sample its probability distribution $P(w)$.  
The method is based on a simple fact that 
remained unnoticed until now: Work on a  
quantum system can be measured by performing 
a generalized quantum measurement at a single time. 
Such measurement, which technically speaking is
denoted as a POVM (positive operator valued 
measure) reduces to an ordinary projective 
measurement on an enlarged system. 
This observation not only demystifies work 
measurement but also suggests a 
new quantum algorithm to efficiently 
sample the distribution $P(w)$. This can be
used, in combination with 
fluctuation theorems, to estimate free 
energies of quantum states on 
a quantum computer. 
\end{abstract}

\maketitle 

\emph{Introduction.--} 
For quantum systems the definition of work 
is rather subtle. As work is not represented 
by a hermitian operator \cite{Talkner07}, it is not an 
ordinary quantum observable. Therefore, work
measurement is certainly atypical. 
It is widely believed that work can only be measured
by performing energy measurements at two 
times \cite{Talkner07,Kurchan00,Tasaki00,Campisi11}.
Instead, here we show that work can be measured  
at a single time 
by means of a very general class of quantum 
measurements which is denoted as a "positive 
operator valued
measure" (or POVM) \cite{Peres,Chuang}. 
This type of generalized measurements  
are standard in quantum optics, quantum metrology,
quantum information, etc \cite{Peres}. In fact, they
define the most general set of questions to which
quantum mechanics can assign probabilities. In
general, they are such that: a) the number of 
outputs may be larger than the 
dimensionality of the space of states and 
b) the states of the system after recording 
different outcomes of the measurement 
are not orthogonal. POVM's can always be realized 
as ordinary projective measurements 
on an enlarged system \cite{Peres,Chuang}. 
Thus, we show that, contrary to the common lore, 
work can be measured at a single time, that its
probability distribution can be efficiently sampled 
and that work is a magnitude with which other 
systems can directly couple. 

Interest on work measurement in classical and 
quantum systems blossomed after the 
discovery of fluctuation theorems, 
the most significant result of statistical 
mechanics in decades 
\cite{Jarzynski97,Crooks99}. Notably, Jarzynski 
identity establishes that for any non--equilibrium
process, the probability $P(w)$ to detect 
work $w$ contains the information required to 
compute free energy differences 
between equilibrium states. This has 
been used to evaluate free energies for classical 
systems at the nano-scale  
\cite{Collin05}. In the quantum regime, 
there have been proposals 
to determine $P(w)$ by measuring energy 
at two times with cold ions  
\cite{Huber08}, to use properties of optical 
spectra to evaluate $P(w)$ \cite{Heyl12}, 
to perform many intermediate measurements 
on smaller  subsystems \cite{Campisi10}, to
adopt alternative strategies for driven two level 
systems \cite{Hekking13}, etc. Recently, the use 
of Ramsey interferometry has been suggested to 
estimate the characteristic 
function of $P(w)$ \cite{Dorner13,Mazzola13}. This
method is based on the well known scattering 
algorithm that estimates the 
average of any unitary operators \cite{Knill98}. 
This was later generalized 
for quantum open systems 
\cite{Campisi14,Watanabe14,Fusco14}
and implemented in 
NMR experiments \cite{Batalhao13}. 

The method we present here
is the only one that directly samples $P(w)$ by
means of a projective measurement at a 
single time.  By virtue of this fact, quantum 
coherence is destroyed only at that final time. 
Until then, the evolution
is unitary. For this reason, this scheme can be used 
to study the role of quantum coherence in 
thermodynamical processes \cite{Lostaglio14,Lostaglio14-2}. 
Our results helps to demystify work 
measurement for
quantum systems. As we show, 
every value of work $w$ can be coherently 
recorded in the state of a quantum 
register (an auxiliary system), which can 
then affect the fate of any other 
system, including the original one. Thus, although 
work is not represented by a hermitian operator, it 
shares the essential properties of standard 
observables. 
Last, but not least, we show that our results 
motivates a novel quantum 
algorithm that,  when executed in a quantum computer, 
would estimate free energies exploiting the efficient 
sampling of $P(w)$. 

The non--existence of a hermitian work--operator
\cite{Talkner07} is a consequence of the 
relation between work and energy differences. 
As the number of possible values of work 
$w=E_f-E_i$ is typically larger than the 
dimension of  the space of states, 
a hermitian operator representing work 
cannot exist. This does not imply that work is 
not measurable. Quite the opposite, work
can be measured using the following 
strategy: Consider a system with  
initial state $\rho(t_0)$, which is driven 
from an initial Hamiltonian $H=H(t_i)$ to a final 
one $\tilde H=H(t_f)$. The results of energy 
measurements at times $t_i$ and $t_f$ are 
eigenvalues of $\tilde H$ and $H$ satisfying  
$H\ket{\phi_n}=E_n\ket{\phi_n}$ 
and $\tilde H|\tilde\phi_m\rangle=\tilde
E_m |\tilde\phi_m\rangle$. In every instance 
work is defined as $w=\tilde E_m-E_n$, 
which is  distributed with probability
\begin{equation}
P(w)=\sum_{n,m}p_n\, p_{m,n}\,\delta(w-(\tilde E_m-E_n)),
\label{eq:workdef}
\end{equation}
where $p_n=\bra{\phi_n}\rho(t_0)\ket{\phi_n}$ is the 
probability to obtain the energy $E_n$ and 
$p_{m,n}=|\bra{\phi_n}U_{f,i}|\tilde\phi_m\rangle|^2$ is
the transition probability between energy 
eigenstates when the system is driven by the 
evolution operator $U_{f,i}=U(t_f,t_0)$. 
From Eq. \eqref{eq:workdef}, we can 
derive the identity $\int dw P(w) \exp(-\beta w)
=\sum_{n,m}p_n\, p_{m,n}\,\exp(-\beta(\tilde E_m-E_n))$. 
For a thermal initial state,   
$\rho(t_0)=\exp(-\beta H)/Z_0$, the 
remarkable identity  derived first in 
\cite{Jarzynski97,Campisi11} follows:
$\langle\exp(-\beta w)\rangle=
\tilde Z/Z=\exp(-\beta\Delta F)$, where $F$ is the 
Helmholtz free energy.

\emph{Work measurement as a generalized measurement.--}  
We can rewrite eq.~\eqref{eq:workdef}
as $P(w)=\tr[\rho W(w)]$ where 
\begin{equation}
W(w)=\sum_{n,m}p_{m,n}\, \delta(w-E_{m,n}) \,
\ketbra{\phi_n}{\phi_n}, 
\label{eq:Ww}
\end{equation}
with $E_{m,n}\equiv(\tilde E_m-E_n)$. Operators 
$W(w)$ define a positive operator
valued measure (POVM) as they form a set
of non--negative operators which 
decompose the identity as $\int dw W(w)=I$.
The operators $W(w)$ are not orthogonal since the 
number of values that $w$ can take 
is larger than dimension of the Hilbert space. 
A POVM defines the most general type of quantum 
measurement one can perform. Neumark's theorem  
\cite{Peres} establishes that any POVM can 
be realized as a projective measurement on an
enlarged system. Applying this observation for the
case of work measurement, we conclude that
it is always possible to design an apparatus such that:
$(i)$ it produces an output $w$ with 
probability $P(w)$; $(ii)$ when $w$ 
is recorded, the system is prepared in a state
$\rho_w$ (that depends on $\rho$, 
$w$, and on the measurement 
implementation). There is not a unique method
to implement a given POVM. Here, we present a 
simple strategy that can be used to evaluate 
work. For this purpose, we can 
couple the system $\sys$ with
an auxiliary system $\aux$ in such a
way that $\aux$ gets entangled with $\sys$
keeping a coherent record of 
the energy at two times. To do this, $\sys$ 
and $\aux$
must interact twice through an entangling 
interaction described by the Hamiltonian 
$H_{I}= \lambda H\otimes\hat p$, where
$\lambda$ is a constant and 
$\hat p$ is the generator of translations 
between the states $\ket{w}$ 
of $\aux$. In the simplest case we can 
consider $\aux$ with a continuous degree 
of freedom, where $\{\ket{w}, \ w\in{\mathcal R}\}$ 
is a basis of its space of states. 
The evolution operator  
$U_{I}=\exp(-iH_{I} t)$ is such that 
\begin{equation}
U_{I}(\ket{\phi_n}\otimes\ket{w=0})=
\ket{\phi_n}\otimes\ket{w=E_n}.
\label{eq:actionofu}
\end{equation} 
Then, we drive the system with the operator $U_E=U_{f,i}=U(t_f,t_i)$. Finally, a new entangling
interaction is applied. In summary, we apply the
unitary sequence 
$U_{IEI}=\tilde U_{I}\ U_E U^\dagger_{I}$
(with $\tilde U_{I}=\exp(-i\lambda \tilde H\otimes \hat p\ t)$). 
The resulting evolution transforms the 
initial product state $\ket{\Psi(t_0)}=
\ket{\phi_0}\otimes\ket{w=0}$ into 
the final entangled state 
\begin{equation}
\ket{\Psi_f}=\sum_{n,m} \langle\tilde\phi_m | U_E\ket{\phi_n}
\bracket{\phi_n}{\phi_0}\;
| \tilde\phi_m \rangle\otimes\ket{w=E_{m,n}}.
\label{eq:psifinal}
\end{equation}
At this stage we measure $\aux$. The probability to 
find $\aux$ in the state $\ket{w}$ is   
$P(w)=\bra{\Psi_f}(I\otimes\ketbra{w}{w})\ket{\Psi_f}$. It is 
simple to show that $P(w)$ is precisely the distribution
given in Eq. \eqref{eq:workdef}.  
The state after detecting work $w$ is 
$\rho_w=A_w\rho A^\dagger_w/P(w)$. Here, 
$A_w$ is such that $W(w)=A^\dagger_wA_w$, 
$P(w)=\tr(\rho W(w))$) and is given as
\begin{equation}
A_w=\sum_{n,m}\delta(w-E_{m,n})
\langle\tilde\phi_m|U_{E}\ket{\phi_n} \;
\ketbra{\tilde\phi_m}{\phi_n}.
\label{eq:Aw}
\end{equation}
Noticeably, contrary to what happens in the standard two-time measurement scheme, the final state $\rho_w$ is \emph{not} 
an eigenstate of the final Hamiltonian.

Thus, we described a  
method to measure work, which is such that 
the outcome $w$ is generated with probability 
$P(w)$, preparing the system in one of the 
non--orthogonal states $\rho_w$. In fact, although
work is not a Hermitian operator, it can be 
measured with an ordinary POVM.  

It is interesting to notice that the 
sequence of operations 
$U_{IEI}=\tilde U_{I}U_EU^\dagger_{I}$ 
has been realized in a recent experiment. The 
interaction $U_{I}$ is precisely the one 
realized in a Stern 
Gerlach (SG) apparatus when the spin ($\sys$)
degrees of freedom interact with the motional ($\aux$)
degrees of freedom of a particle when it enters 
an inhomogeneous magnetic field. Then, 
the momentum of the 
particle is shifted by an amount that 
depends on the projection of the spin along 
the field. The magnitude 
of the shift depends on the field 
gradient and on the interaction time
(controlled by the velocity of the particle). 
To realize $U_{IEI}$ we need a sequence
of two SG apparatus with a spin driving field
in between. Notably, this was 
done in a recent experiment \cite{Folman} 
where SG type interactions were used to create 
coherent superpositions of momentum 
wave packets of an atomic beam. This remarkable
experiment was done using an atom chip manipulating
a falling cloud of $^{87}\textrm{Rb}$ atoms 
obtained from a BEC. The SG interaction $U_{I}$
was implemented using a gradient pulse 
generated by coils in the chip. The gradient acts 
as a beam splitter and, as a consequence the 
atomic cloud splits into two pieces that move with 
different momenta, depending on their internal 
(Zeeman) state. As demonstrated in the experiment
\cite{Folman}, the atoms  behave as two-level 
systems and, after splitting the atomic cloud, the 
coils in the chip can generate radio frequency
pulses coherently driving transitions between 
the Zeeman sub-levels $\ket{F,m_F}=\ket{2,2}$ 
and $\ket{2,1}$. This implement the operator
$U_E$, the second step of the $U_{IEI}$ sequence. 
Finally, as shown in \cite{Folman}, a new $U_{I}$ 
interaction can be applied to split the wave packet 
for a second time. As a result, four atomic clouds are 
produced, whose densities were measured 
by recording the shadow of the atoms in a resonant 
absorption experiment, tuned to an appropriate  
transition. 

Here, we simply stress that a 
recent experiment performed with a 
different purpose  \cite{Folman},  
can be interpreted as 
realization of the work measurement
method presented above. In that case, the initial 
and final Hamiltonians are defined by the  
gradient pulses (and are proportional to the  
interaction times) while the driving field is determined
by the intermediate radio frequency pulses.
 Each of the four spots observed in the final
image correspond to one of the four results of the 
POVM. Thus, the image in \cite{Folman} directly 
reveals the work distribution for a single driven
spin-1/2 particle. Different driving processes can
be easily implemented.

\emph{Work estimation through phase estimation.--}
The above method to measure work 
naturally translates into a quantum 
algorithm that efficiently sample $P(w)$. 
The algorithm would run on a quantum 
computer which could be used to efficiently 
estimate moments of the work distribution. 
The method is a variant of the phase 
estimation algorithm \cite{Chuang}, that plays 
a central role in many quantum algorithms. 
We consider an $N$-qubit system $\sys$ 
($D_\sys=2^N$)
and an $M$-qubit ancilla $\aux$ ($D=2^M$ 
determines the precision of the sampling, as 
described below). 
We assume for simplicity that the 
Hamiltonians $H$ and $\tilde H$
have bounded spectra that take values between   
$\pm E_M/2$ (this condition can be relaxed). 

\begin{figure}[tb]
    \includegraphics[width=0.45\textwidth]{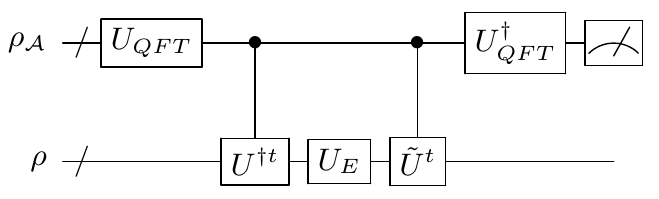}
\caption{Quantum circuit for the estimation of 
work probability distribution. The initial state of the 
auxiliary $M$--qubit system is a pure state $\rho_\aux=\ket{\psi_\aux}\bra{\psi_\aux}$ with $\ket{\psi_\aux}=\ket{x=0}$.
When the initial state of the system is pure given by 
$\ket{\phi}=\sum_{n=1}^{D_\sys}c_n \ket{\phi_n}$ 
then the state of $\sys$ and $\aux$ just before the 
measurement is
$\ket{\Psi}_{\sys,\aux}=\frac{1}{D}
\sum_{n,m=1}^{D_\sys} c_n
\bra{\tilde \phi_m}U_E\ket{\phi_n} 
\sum_{t,x}e^{i\frac{2\pi}{D}t
(x-\frac{E_{m,n}}{4E_M}D)}
\ket{w}\otimes\ket{\tilde\phi_m}$.}
\label{fig:algorithm}
\end{figure}

The algorithm below produces an 
an $m$--bit string output $x$ 
with a probability $P_D(x)$, 
which is a coarse-grained version of the 
work distribution $P(w)$ given in \eqref{eq:workdef}. 
Each integer $x$ identifies a certain amount of 
work through the identity $w=4E_M x/D$. Positive
(negative) values of $w$ correspond to $0<x\le D/4$
($3D/4\le x\le D-1$).
The quantum algorithm for sampling $P(w)$, 
shown in Fig. \ref{fig:algorithm}, has 
six steps: $(i)$ prepare 
the initial state $\ket{x=0}$ for $\aux$  and
$\rho$ for $\sys$; $(ii)$ apply a quantum Fourier 
transform (QFT) 
on $\aux$ mapping $\ket{x}$ onto its
conjugate state $\ket{\tilde x}=U_{QFT}\ket{x}=
\frac{1}{\sqrt{D}}\sum_{t=0}^{D-1}
e^{i\frac{2\pi xt}{D}}\ket{t}$;
$(iii)$ apply the controlled operator 
$U_{I}=\sum_{t=0}^{D-1} \ketbra{t}{t}\otimes
U^{\dagger t}$, where 
$U^t=\exp(-i\pi H \ t/4E_M)$; 
$(iv)$ apply the unitary driving
$U_{f,i}$ over $\sys$; 
$(v)$ apply another controlled operation 
$\tilde U_{I}=\sum_{t=0}^{D-1} \ketbra{t}{t}
\otimes  \tilde U^t$, with 
$\tilde U=\exp(-i\pi \tilde H\ t /4E_M)$;
$(vi)$ apply the inverse QFT in $\aux$ 
and measure its state in the $\ket{x}$ basis. 
The algorithm applies the IEI sequence 
described above since the phase estimation 
subroutine is nothing but a standard 
measurement interaction. 

The probability to detect $x$ 
in $\aux$ is 
$P_D(x)=\sum_{m,n}p_n p_{m,n} |F_D(
4E_M\frac{x}{D}-E_{m,n})|^2$
where
\begin{equation}
\left|F_D(z)\right|^2=\left|\frac{1}{D}\sum_{t=0}^{D-1}  
e^{-i\frac{\pi z}{2E_M}t}\ \right|^2=\frac{1}{D^2}\frac{
\sin^2(\frac{\pi z D}{4E_M})}{\sin^2
(\frac{\pi z}{4E_M})}
\label{eq:filter}
\end{equation}
The distribution $P_D(x)$ is 
a coarse-grained version of the true work 
distribution $P(w)$ given in \eqref{eq:workdef}. 
Thus, $P_D(x)$ is the convolution between 
$P(w)$ and the filter function defined
in \eqref{eq:filter}:
\begin{equation}
P_D(x)=
\int dw' \left|F_D\left(4E_M\frac{x}{D}-w'\right)\right|^2 P(w').
\label{eq:estimwork}
\end{equation}
The operators $A_D(x)$ defining the 
POVM are such that 
$P_D(x)=\tr(\rho A^\dagger_D(x)A_D(x))$. They 
are also a convolution between the 
exact expression \eqref{eq:Aw} and a
filter function, i.e.  
$A_D(x)=\int dw'\ F_D(4E_M\frac{x}{D}-w') A(w')$. 

\begin{figure}[tb]
    \includegraphics[width=0.48\textwidth]{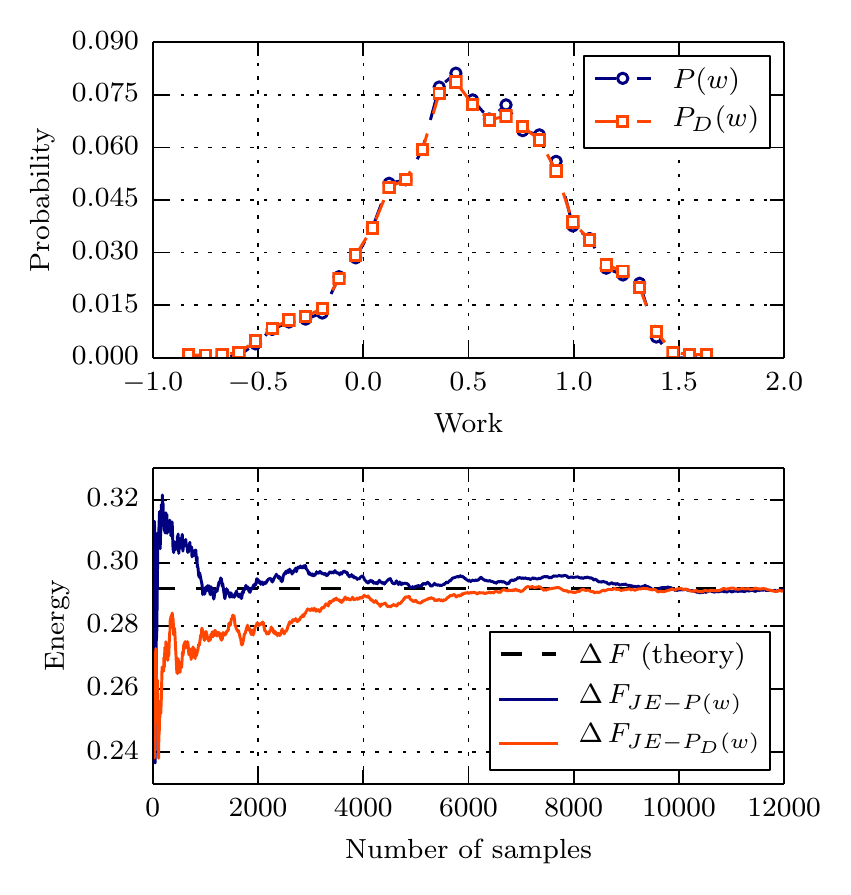}
    \caption{
    Top: Comparison of the coarse-grained version of the exact
    probability distribution (given by \eqref{eq:workdef}) with the
    probability distribution generated by the algorithm (equation
    \eqref{eq:estimwork}). For this example a system $\sys$ of 10-qubits
    was used (therefore giving $2^{20}$ different possibles values of
    work); while the ancilla $\aux$ was composed of only 5-qubits.
    Bottom: Free energy estimation using Jarzynski's equality and work values
    sampled from the exact distribution, $P(w)$, and the distribution
    resulting from the algorithm, $P_D(w)$. It is also shown the exact value
    of the free energy difference, calculated as the ratio of the partition
    functions.
    }
    \label{fig:pcomp_jarz}
\end{figure}

It is simple to show that $P_D(x)$ rapidly 
approaches $P_{cg}(x)$, defined 
as the convolution of $P(w)$ with a rectangular 
function (which is unity for $|w|\le 2E_M/ D$ 
and zero otherwise). 
In fact, if $P(w)$ is bounded, 
then it is straightforward to show that  
$\norminf{P_{cg} - P_D} = 
\mathcal{O}\left(\norminf{P}/D\right)$.
Therefore, 
the difference between $P_D(x)$ and 
$P_{cg}(x)$ decreases exponentially 
with the size of $\aux$.
In Figure \ref{fig:pcomp_jarz} we compare 
$P_{cg}(w)$ and 
$P_D(w)$ in a quenched process between 
two random Hamiltonians $H$ 
and $\tilde H$. As $N=10$, the number of 
different values of $w$ is $2^{20}$.
It is clear that even for a small $\aux$
(with $M=5$ qubits), the sampling 
of the coarse-grained work distribution 
is highly accurate. 


\emph{Estimating the free energy by 
sampling over the work
distribution.--} Sampling the work 
distribution $P(w)$ can be 
useful to efficiently estimate its moments. 
In turn, using Jarzynski 
identity, this can enable the estimation of 
the free energy of quantum states. For this 
one needs the expectation  
$\int dw P(w)\exp(-\beta w)$. 
The above quantum algorithm enables 
sampling the coarse grained distribution 
$P_D(x)$, that can be used to 
efficiently estimate 
averages such as $\langle w\rangle$ 
with an accuracy that  
depends on the number of sampling points,
$K$, as $1/\sqrt{K}$. 
So, for fixed precision (independent of the 
dimensionality of the Hilbert space of $\sys$) 
this method is efficient. In Figure 2 we show the 
dependence of the estimated $\Delta F$ with the 
number of times the distribution $P_D(x)$ is 
sampled (for two random Hamiltonians of a system
of $N=10$ qubits). However, as it is the case
for classical systems, this strategy 
is not always efficient. In fact, efficiency 
depends on the properties of $P(w)$, 
because negative
values of work, for which $\exp(-\beta w)$ is
large are typically under represented in the 
sampling process (a situation that becomes 
worse at low temperatures).

\emph{Summary and comparison with previous work.--} 
We showed that work measurement is a 
generalized quantum measurement (a POVM). 
This observation inspired a new 
method to measure work by performing a 
projective measurement on an enlarged system 
at a single 
time. This method inspires a new 
interpretation of an existing double 
SG experiment  \cite{Folman} and also 
a new quantum algorithm to efficiently 
sample a coarse-grained version of the work 
distribution $P(w)$. This algorithm could run in a 
quantum computer producing an $M$-bit output 
$x$ with a probability $P_D(x)$, which is such 
that $P_D(x)=P(w\in I_x)$ with an accuracy that 
grows exponentially with $M$. Here, $w\in I_x$
iff $\left|w\mp4E_Mx/D\right|\le 2E_M/D$ (where the 
$\mp$ sign respectively corresponds to the cases
$0\le x\le D/4$ and $3D/4\le x\le D-1$). 
It is worth comparing this new method with the 
evaluation the characteristic function 
of $P(w)$ ($\chi(s)$) \cite{Dorner13,Mazzola13}. 
In that case, the estimation of the  
expectation value of a single qubit operator 
is required for each value of $s$. 
By doing this, one can 
efficiently estimate work averages, which 
are obtained from derivatives of $\chi(s)$ at the 
origin. 
However, this method is not efficient 
to sample $P(w)$, which is obtained as the 
Fourier transform $\chi(s)$: 
To achieve the 
same precision we attain using $M$ qubits 
in ${\mathcal A}$, the Ramsey method
\cite{Dorner13,Mazzola13} would need to 
evaluate $\chi(s)$ in $2^M$ points. 
Our method allows the efficient estimation of 
global properties of $P(w)$ (like periodicities) 
and of the free energy for certain families 
of Hamiltonians. 
Finally, we stress that in order to evaluate free 
energies our method requires 
a thermal equilibrium state 
$\rho=\exp(-\beta H)/Z_\beta$ as a resource
(the same as in \cite{Dorner13,Mazzola13}). 
However, this resource is not necessary if we 
use the recently proposed quantum Metropolis 
algorithm that enables
the efficient sampling over the Gibbs ensemble
\cite{Temme11}. 

This work was partially supported by grants from 
Anpcyt (PICT 02843), Conicet and Ubacyt.

\end{document}